\documentclass[final,5p,number,super,times]{elsarticle}
\journal{Seizure}

\usepackage{graphicx}  
\usepackage[english]{babel}
\usepackage{pifont}
\usepackage{amssymb, amsfonts, amsmath, subfigure}  
\usepackage{multirow}  
\usepackage{rotating}  
\usepackage{units}
\usepackage{xspace}
\usepackage{commath}
\usepackage[strings]{underscore}

\newcommand {\bfI} {{\bf I}\xspace}

\newcommand {\cs} {$\mathcal{C}^S$\xspace}
\newcommand {\cc} {$\mathcal{C}^C$\xspace}
\newcommand {\cb} {$\mathcal{C}^B$\xspace}
\newcommand {\ce} {$\mathcal{C}^E$\xspace}

\newcommand {\catf}{\textit{f}\xspace}
\newcommand {\catn}{\textit{n}\xspace}
\newcommand {\cato}{\textit{o}\xspace}

\begin{document}

\begin{frontmatter}

\title{How important is the seizure onset zone for seizure dynamics?}

\author[epi,hiskp]{Christian Geier}\corref{cor1}
\ead{geier@uni-bonn.de}
\author[mpi]{Stephan Bialonski}
\author[epi]{Christian E. Elger}
\author[epi,hiskp,izks]{Klaus Lehnertz}
\address[epi]{Department of Epileptology, University of Bonn, Sigmund-Freud-Stra{\ss}e~25, 53105~Bonn, Germany}
\address[hiskp]{Helmholtz Institute for Radiation and Nuclear Physics, University of Bonn, Nussallee~14--16, 53115~Bonn, Germany}
\address[mpi]{Max Planck Institute for the Physics of Complex Systems, N{\"o}thnizer Stra{\ss}e~38, 01187~Dresden, Germany}
\address[izks]{Interdisciplinary Center for Complex Systems, University of Bonn, Br{\"u}hler Stra{\ss}e~7, 53175~Bonn, Germany}

\cortext[cor1]{Corresponding author}

\begin{abstract}

\noindent
\emph{Purpose:} 
Research into epileptic networks has recently allowed deeper insights into the epileptic process.
Here we investigated the importance of individual network nodes for seizure dynamics.

\noindent
\emph{Methods:}
We analysed intracranial electroencephalographic recordings of 86 focal seizures with different anatomical onset locations.
With time-resolved correlation analyses, we derived a sequence of weighted epileptic networks spanning the pre-ictal, ictal, and post-ictal period, 
and each recording site represents a network node.
We assessed node importance with commonly used centrality indices that take into account different network properties.

\noindent
\emph{Results:} 
A high variability of temporal evolution of node importance was observed, both intra- and interindividually.
Nevertheless, nodes near and far off the 
seizure onset zone (SOZ) were rated as most important for seizure dynamics more often (65~\% of cases) than nodes from within the SOZ (35~\% of cases).

\noindent
\emph{Conclusion:}
Our findings underline the high relevance of brain outside of the SOZ but within the large-scale epileptic network for seizure dynamics.
Knowledge about these network constituents may elucidate targets for individualised therapeutic interventions that aim at preventing seizure generation and spread.

\end{abstract}


\end{frontmatter}

\section{Introduction}
\label{sec:Intro}
Research over the last decade has provided strong evidence for the existence of epileptic (also referred to as epileptogenic) networks comprising cortical and subcortical areas in the genesis and expression of not only primary generalised but also focal onset seizures \cite{Spencer2002,vanDiessen2013,Lehnertz2014}, which has led to new concepts and terminology for classifying seizures and epilepsies \cite{Berg2011}.
A network (or graph) is usually considered as a set of nodes and a set of links, connecting the nodes. Functional (or interaction) brain networks can be derived from measurements of neural activity, and the connectedness  between any pair of brain regions (nodes) can be assessed by evaluating interdependencies between their neural activities.

In addition to investigating structural alterations of epileptic brain networks, studies of functional alterations that make use of electroencephalographic recordings have identified network properties that provide new insights into global aspects of seizure dynamics \cite{Kramer2008,Schindler2008a,Kramer2010,Bialonski2011b,Bialonski2013} and the inter-ictal state \cite{vanDellen2009,Horstmann2010,Kuhnert2010,Kramer2011}.
In the majority of studies, methods from graph theory \cite{Boccaletti2006a,Bullmore2009} had been employed which allow one to characterise global properties such as the clustering in an epileptic network, its efficiency to transport information, or the stability of the globally synchronised state.

There are by now only a few studies that investigated the relevance of local network properties for the dynamics of focal seizures \cite{Kramer2008,Wilke2011,Varotto2012,Zubler2014}.
The \textit{importance} of nodes and links within the network is usually assessed with so-called centrality indices, and each of these indices characterises importance differently by taking into account the diverse roles nodes or links play in a network \cite{Koschutzki2005,Masuda2010,Rubinov2010,Zhang2011c,Klemm2012,Joye2013}.
For patients with seizures arising from neocortex \cite{Wilke2011} or from focal cortical dysplasias \cite{Varotto2012}, most important network nodes have mainly been observed to coincide with the seizure onset zone (SOZ).
These nodes have been interpreted as so-called network hubs that are assumed to play a leading role in the generation and propagation of ictal activity \cite{Wilke2011,Varotto2012}.
These findings, however, may be debated taking into account shortcomings of previous investigations (such as a limited number of seizures, a limited number of investigated brain regions, or usage of only one or a few centrality indices) as well as the many previous studies that reported on the high relevance of brain outside of the SOZ for seizure dynamics \cite{Mormann2005,Kalitzin2005,Kramer2008,Kuhlmann2010,FeldwischStaniek2011,Stamoulis2013,Seyal2014}.

Here, we investigated the importance of nodes in large-scale epileptic networks, derived from a large, heterogeneous set of focal seizures with different anatomical onset locations.
By employing different but commonly used centrality indices \cite{Sporns2007} we aimed at assessing a more comprehensive characterisation of importance of the SOZ, its neighbourhood, and of all other investigated brain regions during the pre-ictal, ictal, and post-ictal period. 
Our findings complement previous studies and extend the understanding on the role of different brain regions in the generation, propagation, and termination of seizures in large-scale epileptic networks.

\section{Methods}
\label{sec:Meth}

\subsection{Clinical data}
The 52 patients (20 women, 32 men; mean age at the time of presurgical evaluation 36~+/-~12 years, range 12~--~65; mean duration of epilepsy 24~+/-~14 years, range 2~--~58) included in this retrospective study suffered from pharmacoresistant focal epilepsy with different anatomical onset locations that required invasive monitoring with intrahippocampal depth electrodes and subdural grid- and strip-electrodes (all manufactured by AD-TECH, WI, USA).
Decisions regarding electrode placement were purely clinically driven and were made independently of this study.
All patients signed informed consent that their clinical data might be used and published for research purposes.
The study protocol had previously been approved by the ethics committee of the University of Bonn.

We analysed intracranial electroencephalographic (iEEG) recordings of 86 epileptic seizures, which were part of previous analyses \cite{Schindler2008a,Schindler2007a}.
They included 38 seizures with mesial-temporal, 22 with extra-mesial temporal, 19 with frontal, 5 with occipital and 2 with parietal lobe onset.
There were 46 complex partial seizures without and 40 with secondary
generalization as judged by studying seizure semiology on the
accompanying video.

Using a Stellate Harmonie recording system (Stellate, Montreal, Canada; amplifiers constructed by Schwarzer GmbH, Munich, Germany) iEEG signals from, on average, 66 electrodes (range 26-124) were band-pass filtered between 0.1-70 Hz, sampled at 200 Hz using a 16 bit A/D converter, and referenced against the average of two electrode contacts outside the focal region.
Reference contacts were chosen independently for each patient.

The peri-ictal recordings lasted, on average, 451~s (range 112~--~1702~s).
The mean seizure duration amounted to 120.2~s (range 33.8~--~395.8~s), with seizure onsets and endings detected fully automatically using the method described by \citet{Schindler2007a}.
We assigned electrode contacts to three location categories, thereby making use of knowledge concerning location and extent of the SOZ, which is defined as the contacts where first ictal discharges were recorded \cite{Rosenow2001}.
Category~\catf ({\em focal}) comprised all contacts located within the SOZ (on average 17.8~\% (2.6~--~52.4) of all contacts over all seizures and contacts) and category~\catn ({\em nearby}) those contacts not more than two contacts distant to those from~\catf (20.5~\% (1.0~--~96.0)).
All remaining contacts were assigned to category~\cato ({\em other}; 61.7~\% (0~--~93.0)).

\subsection{Construction of functional networks}
In order to construct functional networks from iEEG recordings, we associated each electrode contact with a network node and defined functional links between any pair of nodes $i$ and $j$---regardless of their anatomical connectivity---using the cross-correlation function (see \ref{app:matrix}) as a simple and most commonly used measure for interdependence between two signals \cite{Schindler2008a,Bertashius1991}.
iEEG data of each window were normalised to zero mean and unit variance. With a sliding-window approach (2.5~s window duration, 500 sampling points; no overlap) we calculated, for each seizure recording, a sequence of undirected, weighted functional networks spanning the pre-ictal, ictal, and post-ictal period.

\subsection{Assessing node importance with centrality indices} 
Centrality indices (for details of calculation, see \ref{app:centralities}) variously assess importance of individual nodes by considering e.g. a node's connectedness to other parts of the network or by its capability to influence other nodes through short paths.
Degree centrality (or strength centrality (\cs) in case of a weighted network) is defined as the number of links (or the sum of their weights) incident upon a node. 
A node with a high \cs is important since it interacts with many other nodes in the network.
Eigenvector centrality (\ce) recursively determines importance of a node not only on the basis of its links to other nodes, but also with respect to how those other nodes are linked (and so on). 
A node with high \ce is important since it has links to many other nodes that are themselves highly linked and central within the network.
Closeness centrality (\cc) expresses the average geodesic (i. e., shortest path) distance of a node to all other nodes.
A node with high \cc is important since it can reach all other nodes in the network via short paths and may thus exert more direct influence over the nodes.
Betweenness centrality (\cb) is defined as the fraction of shortest paths between pairs of nodes that pass through a given node.
A node with high \cb is important since it connects different regions of the network by acting as a bridge and thus can control the information flow in the network.

The complex spatial and temporal changes in frequency content are known to influence statistical properties of functional networks---such as clustering coefficient, average shortest path length and betweenness centrality---derived from seizure recordings \cite{Bialonski2011b,Wilke2011}. 
In order to avoid spurious centrality estimates, that can trivially be related to spectral properties of the iEEG recording we applied a correction scheme (for details, see \ref{app:surrogatecorrection}), and in the following, we refer to these corrected centrality indices.

\section{Results}

\begin{figure*}
\centering{\includegraphics[width=1.0\textwidth]{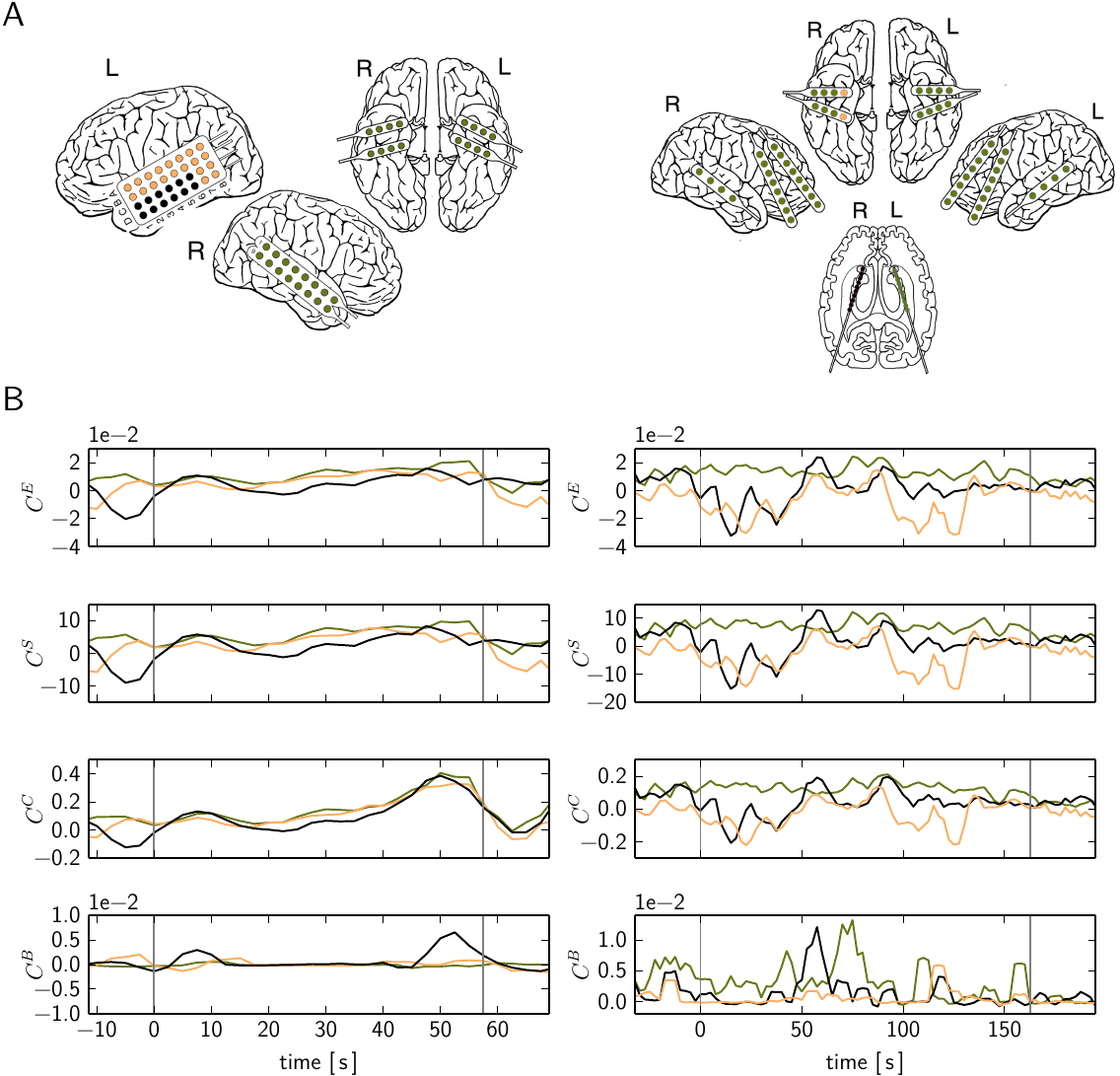}}
\caption{
(A) Schematics of implanted electrodes from a patient with left extra-mesial temporal SOZ (left) and from a patient with right mesial temporal SOZ (right).
Colors indicate location categories to which electrode contacts~(nodes) belong: SOZ~(\catf), black; nearby~(\catn), orange (light grey); other~(\cato), green (dark grey).
(B) Temporal evolutions of centrality indices (top to bottom: eigenvector centrality (\ce), strength centrality (\cs), closeness centrality (\cc), and betweenness centrality (\cb)) of selected nodes from the location categories.  
From each category the node with the highest average centrality during the ictal phase was selected.
Colors as in (A).
The grey vertical lines indicate the beginning and end of the seizure.
For readability, time profiles are smoothed using a moving average (three-point).
}
\label{fig:figure1}
\end{figure*} 

With our analyses we observed a high variability of the various centrality indices for nodes in functional networks spanning the pre-ictal, ictal, and post-ictal period.
In Fig.~\ref{fig:figure1} we show, for each centrality index, temporal evolutions of the centrality values of a node from each of the three location categories for two focal seizures.
From the nodes within each category, we show data from the one with highest average centrality over the course of the seizure.
Interestingly, although the employed centrality indices rated importance of nodes differently, there was a rather close relationship between the temporal evolutions of \cs, \cc, and \ce (Pearson correlation coefficients ranged between 0.85 and 1.00) and these indices rated the same node from each location category as most important (highest respective centrality value). 
In contrast, \cb behaved differently and, with this centrality index, some prominent peaks could be observed for a node from the SOZ (category~\catf) during both seizures.

During the course of the seizures, none of the sampled brain regions was rated as most and constantly important.
We note that neither the described temporal evolutions of node centralities nor some prominent features could be regarded as exemplary for all investigated seizures. 
Nevertheless, the observed relationships between centrality indices were quite stable over all seizures (Pearson correlation coefficients (means and standard deviations); 
(\cs, \ce): $0.99 \pm 0.00$; 
(\cs, \cc): $0.89 \pm 0.18$; 
(\cs, \cb): $0.20 \pm 0.14$; 
(\ce, \cc): $0.87 \pm 0.18$;
(\ce, \cb): $0.18 \pm 0.15$; 
(\cc, \cb): $0.21 \pm 0.15$).
Due to the strong relationships observed for \cs and \ce, we restrict the following presentations to data obtained with \cs, \cc, and \cb.

Because of the high variability of temporal evolutions of node centralities and taking into account the different durations of seizures investigated here, we partitioned each seizure into five equidistant time bins and, in addition, regarded a pre-ictal and a post-ictal bin with a duration that corresponded to a seizure bin \cite{Schindler2008a, Bialonski2011b}. 
For each centrality index, we assigned the time-dependent centrality values to the respective time bins.
In order to control for the different numbers of electrode contacts across location categories, we then determined, for each location category and each time bin, the third quartile of the respective distribution of centrality values.
Eventually, we regarded the category with the highest third quartile value as the most important category for that time bin.

\begin{figure}[htp] \centering{
        \includegraphics[width=\columnwidth]{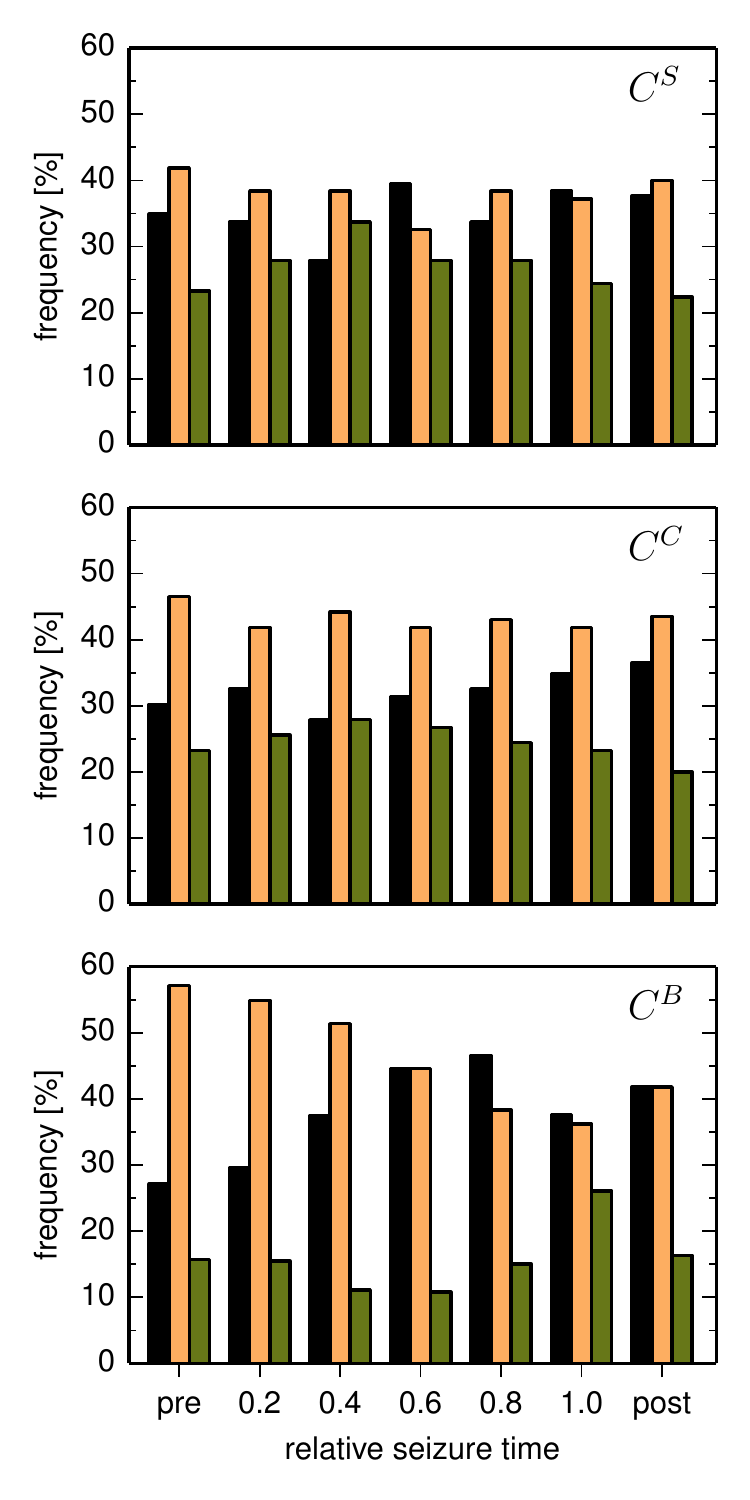}}
\caption{Frequencies with which brain regions (seizure onset zone (\catf), black; nearby (\catn), orange (light grey); other (\cato), green (dark grey)) are indicated as most important for pre-seizure, discretised seizure, and post-seizure time periods using strength (\cs, top), closeness (\cc, middle), and betweenness centrality (\cb, bottom).
Seizures were partitioned into five equidistant time bins.
}
\label{fig:figure2}
\end{figure} 

In Fig.~\ref{fig:figure2}, we show how often which brain region (location category) is indicated as most important over the course of the 86 seizures.
Strength centrality \cs indicated the SOZ (category \catf) and its neighbourhood 
(category \catn) to attain highest importance approximately equally often (in about 30~--~40~\% of cases), with only minor differences as seizures evolved.
Other brain regions (category \cato) were rated most important in only about 20~--~25\% of cases, except during the middle phase of seizures, where importance frequency increased above 30~\%. 

Interestingly, although we observed the relationship between strength \cs and closeness centrality \cc to be quite high, the latter index rated brain regions neighbouring the SOZ (i.e., from category \catn) as most important in about 40~--~45~\% of cases.
Importance frequency of these nodes was rather stable as seizures evolved.
The SOZ (category \catf) and other brain regions (category \cato) were rated as most important only in about 20~--~30~\% of cases, both pre-ictally and up to the middle phase of seizures. 
Towards the end of seizures and extending into the post-ictal phase, importance frequencies of the SOZ and other brain regions exhibited divergent trends, with the former slightly increasing up to 35~\% and the latter slightly decreasing down to 20~\% of cases.

With betweenness centrality \cb, we attained a completely different picture.
Pre-ictally, brain regions neighbouring the SOZ (i.e., from category \catn) were rated most important in more than half the cases (around 55~\%). 
As seizures evolved, the high abundance of these nodes decreased, reaching a minimum importance frequency of about 35\% at seizure ending and then slightly increased again to about 40\% post-ictally.
In contrast, in only about 25\% of cases was the SOZ (category \catf) rated most important pre-ictally, but importance frequency increased up to 45\% towards the last quarter phase of the seizures. 
At the end of the seizures and extending into the post-ictal period, importances of the SOZ and its neighbourhood were rated with around the same frequency.
In only 10~--~15\% of cases were other brain areas (category \cato) rated as most important, except for the end of seizures, where importance frequency increased up to 25\%.

\section{Discussion}
We investigated the importance of different brain regions in large-scale epileptic networks derived from multichannel iEEG recordings for the generation, propagation, and termination of 86 focal seizures with different anatomical onset locations.
Importance of network nodes can be characterised with various centrality indices \cite{Koschutzki2005,Masuda2010,Rubinov2010,Zhang2011c,Klemm2012,Joye2013} but it is not yet clear, which index is best suited for a characterisation of peri-ictal network dynamics.
We therefore decided to employ indices that had been used most often in other network studies, namely strength (or degree), closeness, betweenness, and eigenvector centrality.  
Previous studies that also investigated the importance of brain regions in large-scale epileptic networks employed at least two indices \cite{Kramer2008,Varotto2012}.

This study revealed three main findings. 
First we observed a high temporal variability of node importance in epileptic networks spanning the pre-ictal, ictal, and post-ictal period.
Second, in about 65\% of seizures, nodes off the clinically defined SOZ were identified---on average---as most important throughout the course of the seizure, while nodes from the SOZ were indicated as important in only 35\% of cases. 
Third, we observed rather strong correlations between strength, closeness, and eigenvector centrality, while betweenness centrality behaved differently from the other three indices.

\subsection*{Temporal variability of centrality indices}
The temporal variability of centrality indices in epileptic networks spanning the pre-ictal, ictal, and post-ictal period was high both inter- and intraindividually. 
We investigated a number of potentially influencing factors such as seizure type (with and without secondary generalisation), the vigilance states seizures arose off, and the anatomical location of the SOZ (data not shown). 
None of these factors appeared indicative of the temporal variability.
Moreover, different seizures from the same patient sometimes yielded similar temporal evolutions of centrality indices and sometimes very different ones.
We also checked a possible impact of some crucial steps of analysis (such as normalisation of the interaction matrix (see \ref{app:matrix}) or the surrogate correction) but the temporal variability was conserved even without these steps.

Yet our findings are in line with previous observations by Kramer et al. \cite{Kramer2008}, who reported on similar temporal evolutions of centrality indices during seizures from four patients.
The high temporal variability seen intra- and interindividually possibly points to crucial but as yet only poorly understood spatial and temporal aspects of seizures. 
These aspects may not be fully identifiable with analysis techniques that characterise seizure dynamics only locally \cite{Jouny2007} or through global, large-scale interactions \cite{Schindler2007a,Schindler2010} but may be better assessable with techniques that take into account local properties within the context of an interaction network \cite{Schindler2012,Bialonski2013}.

\subsection*{Importance of the seizure onset zone}
In contrast to previous studies \cite{Wilke2011,Varotto2012}, which reported most important network nodes to coincide with the seizure onset zone, our investigations indicated the latter to be, in general, neither more nor less often important than other nodes in epileptic networks spanning the pre-ictal, ictal, and post-ictal period.
This discrepancy might be due to the applied methodology to derive network links (directed vs. weighted), the higher number of seizures from a higher number of patients investigated here as well as due to a higher number of recording sites which leads to networks of vastly increased size (i.e., number of nodes).
Moreover, we considered focal seizures with different anatomical onset locations.

Our findings are, however, in line with previous reports on the high relevance of brain outside of the SOZ but within the epileptic network for seizure dynamics \cite{Mormann2005,Kalitzin2005,Kramer2008,Kuhlmann2010,FeldwischStaniek2011,Stamoulis2013,Seyal2014}. 
Particularly brain areas neighbouring the SOZ were most often rated as important (using betweenness centrality) pre-ictally and during the first half of a seizure which would characterize these non-focal nearby structures as a bridge between the SOZ and other brain regions. 
This would support previous reports on a \emph{decoupling} of the SOZ from the rest of the brain that has been observed interictally \cite{Warren2010,Geier2013,Lehnertz2014}, pre-ictally \cite{Arnhold1999,Mormann2000,Chavez2003b,Mormann2003b}, and at seizure onset \cite{Netoff2002,Wendling2003,Cymerblit2012}.
One might speculate, whether network nodes that were identified as most important for seizure dynamics but were located far off the SOZ (on average, in 23\% of cases) could serve as target for resective therapies, particularly in cases where the SOZ is located within or close to eloquent cortex and thus can not be accessed surgically (see, however, \citet{Schramm2008}).  
Moreover, these nodes and in particular brain areas neighbouring the SOZ might also serve as target for novel therapeutic intervention in order to prevent or abort ictal activities \cite{Fisher2012,Wu2013}.
Nodes neighbouring the SOZ were rated as most important pre-ictally in up to 60\% of cases, and we might thus hypothesise that they not only facilitate seizure generation but may be a better target for prevention strategies.
For these alternative therapy options to become feasible, however, methodological improvement as well as prospective studies are needed, including studies on the importance of network nodes during the interictal state. 
On the other hand, since importance frequency of the SOZ increased towards the end of the seizures, we might further hypothesise that this brain region plays a role not only in seizure spread but also in seizure termination. 
A better understanding of large-scale interactions underlying seizure dynamics in epileptic networks may elucidate targets for treatments that aim at preventing or at least confining seizure spread, which has devastating consequences for patient safety and quality of life \cite{Surges2012}.

\subsection*{Similarities and differences between centrality indices}
Although the centrality indices employed here rate node importance differently, we observed a very strong correlation between strength and eigenvector centrality, and to a lesser extent also between strength and closeness centrality.
This finding is in line with previous studies that reported on similar correlations, although for networks of different origin \cite{Estrada2010b,Kuhnert2012}.
Betweenness centrality behaved differently from the other indices, although betweenness and closeness centrality rely on the concept of shortest paths.
Betweenness centrality identifies nodes as most important that are \emph{between} most other network nodes, but it remains to be shown whether nodes identified as important with this index indeed facilitate seizure dynamics.
Given the fact that a number of electrode contacts usually comprise the SOZ, it also remains to be shown whether other centrality indices are better suited to identify important nodes in epileptic networks.

\subsection*{Limitations of the study}
With intracranial recordings, access to brain regions other than those suspected to be involved in the epileptic process is limited, thus undersampling bias is inevitable.
Since electrode placements were driven by clinical needs in each patient and were thus not standardised, an electroencephalographic signal is not representative of exactly the same anatomic regions in each patient.
In addition, different number of electrodes lead to networks of different size, and it is not yet clear how exactly this affects global and local network indices and how to compare such networks \cite{Bialonski2010,Joudaki2012}. 
Our patients received different antiepileptic drugs (AED) with different mechanisms of action, and the majority of patients were under combination therapy with two or more AED. 
It is, however, not known if and to what extent AED affect the global and local properties of epileptic networks.

\section{Conclusion}
In summary, our study suggests that in only a limited number of cases, the SOZ can be regarded important for the generation, propagation, and termination of seizures.
Monitoring the importance of other brain regions that together with the SOZ constitute the epileptic network, can help to identify network nodes, which are crucial for seizure facilitation and termination and can thus be regarded as potential targets for individualised focal therapies.

\section{Acknowledgements}
We thank Gerrit Ansmann, Henning Dickten, and Stephan Porz for helpful comments on earlier versions of this manuscript. 
This work was supported by the Deutsche Forschungsgemeinschaft (Grant No. LE660/4-2). 

\section{Disclosure}
None of the authors has any conflict of interest to disclose.

\appendix

\section{Interaction matrix}
\label{app:matrix}
Given iEEG signals $x_i(t)$ and $x_j(t)$ from electrodes $i$ and $j$ ($i,j = 1, \ldots,N$), normalised such that each has zero mean and unit variance, the normalised maximum-lag correlation reads

\begin{equation}
    \label{eq:rho}
    I_{ij} = \max_{\tau} \left\{  \abs{\frac{K(x_i, x_i)(\tau)}{\sqrt{K(x_i, x_i)(0) K(x_j, x_j)(0)}}}  \right\},
\end{equation}
with the cross-correlation function
\begin{equation}
K(x_i,x_j)(\tau)= 
	\left\{
		\begin{array}{ll}
			\sum_{t=1}^{T-\tau} x_i(t+\tau)x_j(t), & \tau \geq 0 \\
			K(x_j,x_i)(-\tau), & \tau < 0
		\end{array}
	\right.
        .
\end{equation}

This function yields high values for such time lags $\tau$ for which iEEG signals $x_i(t+\tau)$ and $x_j(t)$ have a similar course in time.  
Calculating $I_{ij}$ for all pairs $(i,j)$ of electrodes, we derive a symmetric weighted interaction matrix \bfI with entries $I_{ij}$ and size $N\times N$, which is usually interpreted as an undirected weighted network~\cite{Bullmore2009}. 
For $i=j$ we set $I_{ij}=0$ to avoid self-connections.
In order to rule out a possible influence of the mean strength of interaction\cite{Horstmann2010,Ansmann2012}, we normalise the interaction matrix \bfI such that it represents a weighted network with a mean weight of 1, by dividing each element $I_{ij}$ by the mean weight of \bfI.

\section{Centralities}
\label{app:centralities}
\noindent
Strength centrality \cs of node $i$ is defined as\cite{Barrat2004b}
\begin{equation}
    \mathcal{C}^S (i) = \frac{\sum_j I_{ij}}{N-1}.
\end{equation}
It can be regarded as a weighted version of degree centrality, which is not a sensible measure for a weighted network in which all links exist (but might have a small weight).

\vspace{0.5cm}
\noindent
Closeness centrality \cc of node $i$ is defined as:
\begin{equation}
    \mathcal{C}^C (i) = \frac{N-1}{\sum_j d_{ij}},
\end{equation}
where $d_{ij}$ is the length of the shortest path between nodes $i$ and $j$. 
On a weighted network, paths can be defined by assuming the length $d_{ij}$ to vary inversely with its weight $I_{ij}$ \cite{Horstmann2010,Wilke2011}.

\vspace{0.5cm}
\noindent
Betweenness centrality \cb of node $i$ is defined as:
\begin{equation}
    \mathcal{C}^B (i) = \frac{2}{(N-1)(N-2)}\sum_{\substack{h=0\\ h \neq i,j}}^{N}\sum_{\substack{j=0\\ j\neq i}}^{N}\frac{\eta_{hj}(i)}{\eta_{hj}},
\end{equation}
with $\eta_{hj}$ the number of all shortest paths between the nodes $h$ and $j$ and $\eta_{hj}(i)$ is the number of these paths running through node $i$. 
We used the algorithm proposed by \citet{Brandes2001} to estimate \cc and \cb.

\vspace{0.5cm}
\noindent
Eigenvector centrality \ce of node $i$ is defined as the $i$th entry of the eigenvector $\vec{v}$ ($v(i)$) corresponding to the dominant eigenvalue $\lambda_{\rm max}$ of the weighted interaction matrix \bfI:
\begin{equation}
\mathcal{C}^E (i) = v(i), 
\end{equation}
which we derive from the eigenvector equation ${\bf I} \vec{v} = \lambda \vec{v}$ using the power iteration method.

\section{Surrogate Correction}
\label{app:surrogatecorrection}
Not taking into account spectral properties of the iEEG recording can result in spurious centrality estimates.
In order to minimise the influence of the power spectrum we applied a surrogate correction proposed by \citet{Bialonski2011b} and by \citet{Bialonski2013}.
To this end, we generated, for each iEEG recording from each electrode contact, twenty surrogate time series \cite{Schreiber1996a}, which have power spectral contents and amplitude distributions that are practically indistinguishable from those of iEEG recording but are otherwise random. 
Amplitudes are iteratively permuted while the power spectrum of each iEEG recording is approximately preserved. 
This randomization scheme destroys any significant linear or non-linear dependencies between iEEG recordings.
Eventually, we performed the same steps of analysis (construction of functional networks and calculation of centrality indices) as described above. 
For all nodes in each network, we corrected the centrality indices by subtracting the respective mean values derived from the surrogate analyses.

\bibliographystyle{elsarticle-num-names}

\end{document}